\documentclass[doublecol]{epl2} 

\usepackage{amsmath}
\usepackage{graphicx}
\usepackage{amsfonts}
\usepackage{amssymb}

\title{A diffusion-induced transition in the phase separation\\ 
  of binary fluid mixtures subjected to a temperature ramp}

\shorttitle{A diffusion-induced transition in phase separation}

\author{Izabella J. Benczik \and J\"urgen Vollmer}
\institute{Max Planck Institute for Dynamics and Selforganization,
  G\"ottingen, Germany}


\abstract{Demixing of binary fluids subjected to slow temperature
  ramps shows repeated waves of nucleation which arise as a
  consequence of the competition between generation of supersaturation
  by the temperature ramp and relaxation of supersaturation by
  diffusive transport and flow.  Here, we use an
  advection-reaction-diffusion model to study the oscillations in the
  weak- and strong-diffusion regime.  There is a sharp transition
  between the two regimes, which can only be understood based on the
  probability distribution function of the composition rather than in
  terms of the average composition.  We argue that this transition
  might be responsible for some yet unclear features of experiments,
  like the appearance of secondary oscillations and bimodal droplet
  size distributions.}

\begin{document}
\maketitle

\section{Introduction}

Liquid-liquid phase separation occurs whenever an isotropic binary
mixture is transferred into a bi-phasic region where its isotropic state is
no longer stable, and the mixture decomposes into two equilibrium
phases. In classical approaches \cite{phaseSep}
demixing was induced by sudden temperature quenches, after which the phase
separation was monitored in constant temperature conditions. In many
industrial \cite{KS92} and natural \cite{S03,S93} applications however,
temperature is not constant but it {\em coevolves} with the phase
separation. Accordingly, some recent works focused on systems
subjected to time-dependent variations of the temperature, like repeated cycles
of cooling and heating \cite{x10}, or slow continuous temperature
ramps \cite{VSV97,AVV05,VVS97,CVWV03,VAV07,LRVH11,V08,BV10}.

Experiments \cite{VSV97,AVV05,LRVH11} in fluid mixtures
subjected to slow temperature ramps show unexpected phenomena:
rather than continuously following the gradual change in temperature, 
the phase separation exhibits consecutive bursts of droplet nucleation 
alternating with quiescent
periods \cite{VSV97,AVV05,VVS97,VAV07,CVWV03,LRVH11,V08,BV10}.

This intriguing phenomenon can be explained in terms of the competition between the 
temperature ramp and diffusion. 
Due to the temperature ramp there are droplets present in the system 
at any time.  Any change of temperature results in a change of the equilibrium 
composition of the background fluid and of the droplets:
\emph{supersaturation} -- defined as the deviation from the actual to the 
equilibrium composition---builds up in the sample. This supersaturation is 
relaxed by diffusive exchange of mass between the droplets and their environment, 
or occasionally also by nucleation of new droplets.

Immediately after a wave of nucleation, the droplet concentration is
high \cite{LRVH11}. In this case, diffusion, which acts on length
scales of typical droplet distances, relaxes supersaturation by orders
of magnitude faster than it is generated by the temperature ramp
\cite{V08}. The concentration remains close to equilibrium, and no new
droplets are nucleated. The system enters a quiescent period.  In the
course of time, the number of droplets in the system decays due to
coarsening of the droplet distribution and sedimentation. Droplet
distances grow. Thus the diffusive exchange becomes slow, and
supersaturation rises again.  Eventually, a new wave of nucleation is
triggered.  Clearly the strength of diffusive transport has a critical
impact on the alternation of the active and quiescent periods of phase
separation.

The strength of the diffusive transport is characterized by the ratio $D/L^2$. Hence, it is 
affected not only by the distance between droplets $L$, but also by the 
diffusion coefficient $D$. The latter effect is especially important 
because the temperature ramp, that drives the phase separation, causes 
considerable variations of $D$. 
It rises monotonously 
as function of the distance from the critical point (cf.~\cite{materials}).

Here, our aim is to investigate systematically the behaviour of the oscillations
in different (weak and strong) diffusion regimes. 
We adopt a numerical approach built on previous mean-field discussions
\cite{VVS97,VAV07} as far as the evolution of the supersaturation
is concerned, and  combined with a reactive-flow
description \cite{BV10}. The latter deals explicitly with the spatial distribution 
of droplets and supersaturation, and with the dynamics of droplets 
(formation, coagulation and advection).
Upon increasing the diffusion coefficient we find a clearly defined transition 
in the characteristics of the oscillations --- in particular in their period.
We will compare this behaviour to experimental results, and point out that some yet 
unclear features of the experiments might be attributed to the continuous growth 
of the diffusion coefficient as the system departs from the critical point.

\begin{figure*}
{\hfill \includegraphics[width=1\textwidth]{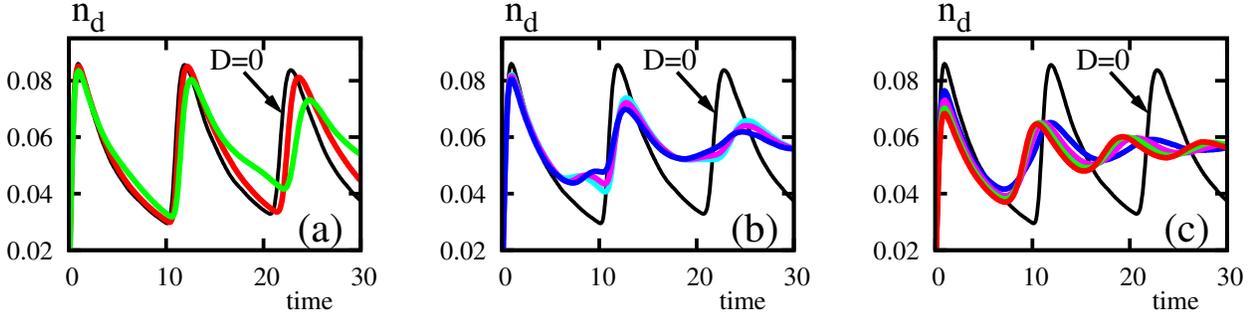}}
\caption {Evolution of the droplet density $n_d$ in numerical simulations
for flow amplitude $A=0.8$ and cooling rate $\xi=0.04$ in the
(a) weak ($D=0.05$ - red, $D=0.1$ - green line), 
(b) intermediate ($D=0.14, 0.15$ and $0.16$ in decreasing order of the amplitudes), and 
(c) strong  ($D=0.25, 0.35, 0.45$, and $0.55$ in decreasing order of the periods) diffusion regimes.
All simulations were started with a uniform initial composition field $\sigma_0
=0.667$. The black thin line represents the diffusionless case. For intermediate 
values of $D$, as shown in panel (b), secondary oscillations appear.}\label{1}
\end{figure*}

\section{The model} 

It is convenient to describe the dynamics in terms of a normalized
composition $\sigma$ \cite{VAV07,V08,BV10}, defined in such a way that
it takes the value $\sigma_0 = 1$ in equilibrium, and values $0 <
\sigma < 1$ when the system departs from equilibrium. In experiments,
the normalized composition decreases like $\dot \sigma = - \xi \sigma$, 
where the decay rate $\xi$ is a function of temperature
\cite{CVWV03,AVV05}. It can be kept almost constant, however, by
choosing a temperature protocol that assures that the equilibrium flux onto the
droplets is constant in time \cite{AVV05,VAV07}.

In order to focus on the spatial fluctuations of the normalized composition we 
formulate the demixing dynamics in terms of the scalar field $\sigma(x,t)$. It is 
subjected to {\em advection} by the background flow, to an exponentially decaying
{\em reaction} 
due to the temperature ramp, and to {\em diffusion} 
(for details see \cite{BV10}):
\begin{equation}
\frac{\partial \sigma(x,t)}{\partial t} + {\bf u} \cdot \nabla \sigma(x,t)
=\Gamma(\sigma) - \xi \sigma(x,t) + \mathcal{D} \Delta \sigma(x,t) \,. 
\label{demix}
\end{equation}
Here $\mathcal{D}$ is the diffusion coefficient, and ${\bf u}$ the
velocity field of the background convection.
The term $\Gamma(\sigma)$ represents formally the nucleation of
droplets (see Eq.~\eqref{pdisc} for the actual implementation as a stochastic process). 
Nucleation brings the system back to equilibrium by
resetting the local composition to $\sigma_0=1$. Thus it brings about 
a non-trivial (i.e. non-zero) final distribution of the
$\sigma(x,t)$ field.

For numerical investigations of the advection-reaction-diffusion equation, Eq.~\eqref{demix},
we use a lattice model proposed and described in an earlier work \cite{BV10}. Here we just outline 
the basic elements of the model.

Spatial degrees of freedom are coarse grained in the form of cells of
size $\varepsilon$ which are labelled by the index $(i,j)$ on an $N
\times N$ lattice defined in a square of size $L$ with periodic boundary conditions.
At each time $\tau$, we apply a reaction step in each lattice cell $(i,j)$,
\begin{equation}
  \sigma(i,j) \to \sigma(i,j) - \xi \; \tau  \; \sigma(i,j), \label{reaction}
\end{equation}
that describes the changes of the normalized composition due to the external cooling. 
This step is followed by a diffusion step according to the transformation (see \cite{P00}):
\begin{equation}
\sigma(i, j) \to \sigma (i,j) + D \; \tau \cdot D(i, j ),
\end{equation} 
where 
$ D(i,j) 
=  -  \sigma(i,j) 
+ [ \sigma(i+1,j) + \sigma(i-1,j) + \sigma(i,j+1) + \sigma(i,j-1) ] /4$ 
is the discrete Laplace operator, and $D$ the diffusion coefficient in the 
discrete map.
In the 
time continuous limit ($\tau \to 0$), this $D$ corresponds to a diffusion coefficient
$\mathcal{D}=D \varepsilon^2/4$.

In the next step, we nucleate droplets in cells that are far from
equilibrium, (i.e. $\sigma<\sigma_{th}\equiv 2/3$)
with a piecewise-linear probability $a(\sigma)$ of nucleation (as in \cite{BV10}):
\begin{equation}
a(\sigma)=
\begin{cases}
0,  & \text{if} \quad \sigma \geq \sigma_{th} = 2/3 \\
2-3 \sigma,  & \text{if} \quad 1/3 \leq \sigma < 2/3 \\
1,   & \text{if} \quad \sigma \leq \sigma_{sp} = 1/3 .
\end{cases}
\label{pdisc}
\end{equation}
This choice reflects on one hand that for $\sigma \leq \sigma_{sp}$
spinodal decomposition causes instantaneous phase separation
\cite{phaseSep,V08}, and on other hand that above a threshold value
$\sigma_{th}$, the mixture is so close to equilibrium that
practically no nucleation events occur. Upon nucleation the
composition is set back to $\sigma_0(x,y)=1$ in each cell where
nucleation took place, as well as in the eight neighbouring cells.

The effects of the background convection are represented by an advection step that 
mixes the lattice cells and droplets in the lattice. This is generated by an 
incompressible alternating shear 
flow (as in \cite{P94}) that produces chaotic trajectories. 
The time-periodic velocity field of the flow is given by
\begin{equation}
\begin{array}{rcl}
   u_x(x,y,t) 
   &=& 
   \frac{AL}{T} \; \Theta\!\left( \frac{T}{2} -t \bmod T \right) \;
   \sin\frac{2 \pi y}{L}, 
   \\[2mm]
   u_y(x,y,t) 
   &=& 
   \frac{AL}{T} \; \Theta\!\left( t \bmod T - \frac{T}{2} \right) \;
   \sin\frac{2 \pi x}{L},
\end{array}
\end{equation}
where $\Theta(x)$ is the Heaviside step function, $T$ the period of
the flow, and $A$ the strength of
the flow.  

Finally, we take into account the merging of droplets: whenever two droplets 
approach each other to within a
distance $r_0$, the two droplets are replaced by a new
droplet placed in the centre of mass of the two droplets.

Henceforth, we fix the cell size to the specific value $\varepsilon^*$ that represents
the maximal spatial scale where the $\sigma$-field takes constant
values in the presence of diffusion and flow.
Below the scale $\varepsilon^*$ the small-scale
fluctuations in $\sigma$, that are created continuously by the straining action
of the velocity field, are dissipated by diffusion \cite{T00}.  
For a flow with Lyapunov exponent $\lambda$ the scale $\varepsilon^*$ can be 
estimated as follows  \cite{T00}. The stretching and folding caused by the chaotic 
flow generates inhomogeneities of the $\sigma$-field on scales that decrease 
according to $\dot \varepsilon = -\lambda \varepsilon$. On the other hand, 
diffusion causes a broadening of the homogeneous areas according to  
$\dot \varepsilon = D/\varepsilon$ (adopting the solution $\varepsilon \sim  t^{1/2}$). 
Thus the size of homogeneous areas evolves as: 
$\dot \varepsilon = D/\varepsilon-\lambda \varepsilon$. This equation possesses a 
steady state solution $\varepsilon^*=(D/\lambda)^{1/2}=L Pe^{-1/2}$, where $\textrm{Pe} 
= \lambda L^2/D$ is the P\`eclet number, a measure of the strength of the diffusivity. 
Thus, the dimensionless cell size amounts to $\varepsilon^* = L \textrm{Pe}^{-1/2}$.
In what follows, we set the length and time scales to $L=1$ and $T=1$.

\section{Oscillations in the weak-  and strong-diffusion regimes}

To explore how the increasing diffusion coefficient modifies the
oscillations of the droplet density we fix the flow rate, the decay
rate of the composition and the droplet interaction radius to $A=0.8$,
$\xi=0.04$, and $r_0=\varepsilon^*$, respectively, and study the
oscillations for different diffusion coefficients ranging from
$D=0.05$ to $D=0.55$. We follow $100$ individual realizations of the
system and average the results.

The oscillations are qualitatively different for low,
intermediate and high values of the diffusion coefficient. In the
weak-diffusion regime, Fig.~\ref{1}(a), the period of the
oscillations increases with increasing $D$, while their amplitude
decreases faster than in the diffusionless case. 
As the diffusion coefficient increases further, this tendency stops at
a certain value of $D$.
For intermediate values of $D$ [Fig.~\ref{1}(b)] secondary
oscillations appear.
Subsequently, the period of the oscillations becomes smaller again
[Fig.~\ref{1}(c)], and the decay rate of the amplitude decreases with
increasing $D$.

\begin{figure*}
{\hfill \includegraphics[width=1\textwidth]{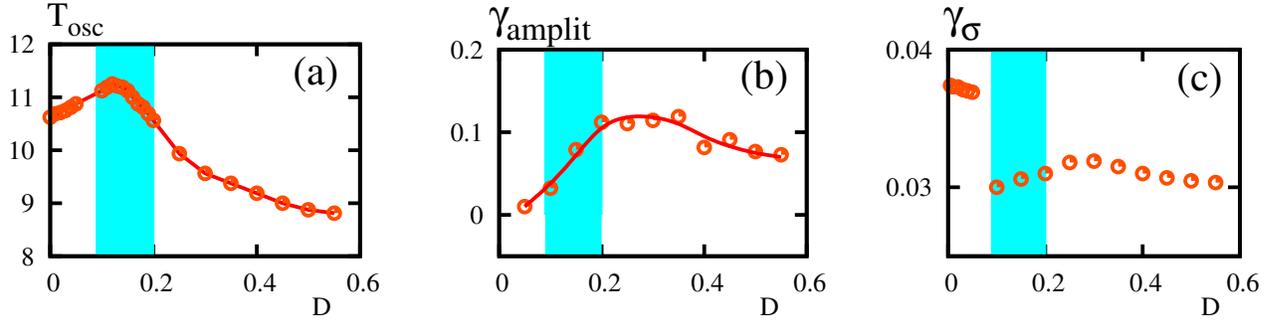}}
\caption {Diffusion induced transition in the oscillatory behaviour of the phase 
separating mixture. For $\xi=0.04$ and $A=0.8$ the 
transition occurs in the highlighted region: $0.1 < D < 0.2$, 
determined as the region where secondary oscillations are 
observable. In the transition region:
(a) the oscillations' period reaches its maximal value, 
(b) the decay rate $\gamma_{\textrm{amplit}}$ of the amplitude reaches its 
maximum, and
(c) the decay rate $\gamma_\sigma$ of the average composition
$\sigma_{\textrm{av}}$ decreases suddenly from $\gamma_\sigma \approx
\xi=0.04$ to $\gamma_\sigma \approx 0.03$. 
}\label{2}
\end{figure*}

\section{Transition between the weak-  and strong-diffusion regimes}

The period of the oscillations takes its maximum value approximately
in the middle of the range where secondary oscillations are observed
[Fig.~\ref{2}(a)].  Furthermore, the appearance of secondary
oscillations in the intermediate diffusion regime is accompanied also
by dramatic quantitative changes of the decay rate of the oscillation
amplitude and of the decay of the composition in the quiescent period
of the oscillations.

We quantify the decay of the oscillation amplitude by following how
the difference between the maximum droplet density and its average
value $ \Delta n_d \equiv n_{d,\textrm{max}} - n_{d,\textrm{av}}$
evolves in time. our data show that for each diffusion coefficient 
this amplitude exhibits an exponential
decay with rate $\gamma_{\mathrm{amplit}}$,
\begin{equation}
  \Delta n_d \sim \exp \left[-\gamma_{\mathrm{amplit}}(D) \cdot t \right] \, .
\end{equation}
As a function of the diffusion coefficient $D$ [Fig.~\ref{2}(b)], this
decay rate has an extremum localized at the end of the
diffusion range in which secondary oscillations are present. The
fastest decay of the amplitude corresponds to that value of the
diffusion coefficient at which secondary oscillations disappear from
the system. For higher diffusion coefficients, the exponent
$\gamma_{\textrm{amplit}}$ decreases slightly. Then it seems to saturate.

The most abrupt change can be observed in the behaviour of the average
composition in the system. We observe that in the quiescent period of
each oscillation the average composition still (cf.~\cite{BV10}) decreases exponentially as
\begin{equation}
  \sigma_{\textrm{av}} \sim \exp\left[ -\gamma_\sigma(D) \cdot t \right] \, ,
\end{equation}
due to the change of temperature. 
In the absence of droplets this decay 
is determined by the local decay, $\sigma(x) \sim \exp(-\xi \cdot t)$, 
of the composition in individual cells, such that  $\gamma_\sigma =\xi$.
When droplets are present in the system
the diffusion of supersaturation into the droplets slows down the decay of the 
average composition. For small values of the diffusion coefficient the magnitude of 
this effect is small, and the average composition still decays with a rate
$\gamma_\sigma \approx \xi$. 
However, when the diffusivity is increased to values where 
secondary oscillation appear, the decay rate of the 
average composition abruptly drops to a much lower value [Fig.~\ref{2}(c)]. 

We conclude that the region where secondary oscillations are present
represents a transition regime between the weak- and strong-diffusion
regimes, where the transition point $D_{\textrm{tr}}$ is defined by
the maximal value of the oscillation period. To determine the
dependence of the transition point on the different parameters of the
system, we run numerical simulations for different decay rates $\xi$
of the composition, and for different flow rates $A$.  The
position of the transition point is not affected significantly by the
flow, but it is very sensitive to the decay rate $\xi$. More
precisely, the transition occurs around 
$D_{\textrm{tr}} \approx 0.16$ for decay rate $\xi=0.04$, 
at $D_{\textrm{tr}} \approx 0.08$ for $\xi=0.02$, and 
around $D_{\textrm{tr}} \approx 0.04$ for
$\xi=0.01$. This is compatible with a linear scaling
\begin{equation}
D_{\textrm{tr}} \simeq  4 \xi.
\end{equation}
The diffusion coefficient $D$ and the decay rate $\xi$
define the lenght scale $\Lambda = \sqrt{D/\xi}$ at which the effects of the temperature ramp
and the effects of diffusion are comparable.  Diffusion can efficiently 
relax the supersaturation accumulated by the continuous temperature
ramp up to a length scale of order $\Lambda$. 
In the continuum limit we find for the transition from the weak to the strong diffusion 
\begin{equation}
  \Lambda_{\textrm{tr}}= \sqrt{\frac{\mathcal{D}_{\textrm{tr}}}{\xi}} 
= \sqrt{\frac{D_{\textrm{tr}} \; \varepsilon^{*2}}{4 \; \xi}}
= \sqrt{\frac{0.16}{4 \cdot 0.04}} \; \varepsilon^* =\varepsilon^*.
\end{equation}
Hence, the transition occurs when the range in which diffusion can
relax the accumulated supersaturation equals the length scale
$\varepsilon^*= L/\textrm{Pe}^{1/2}$ where the effects of mixing by advection and diffusion
are of the same order of magnitude.

\begin{figure*}
{\hfill \includegraphics[width=0.95\textwidth]{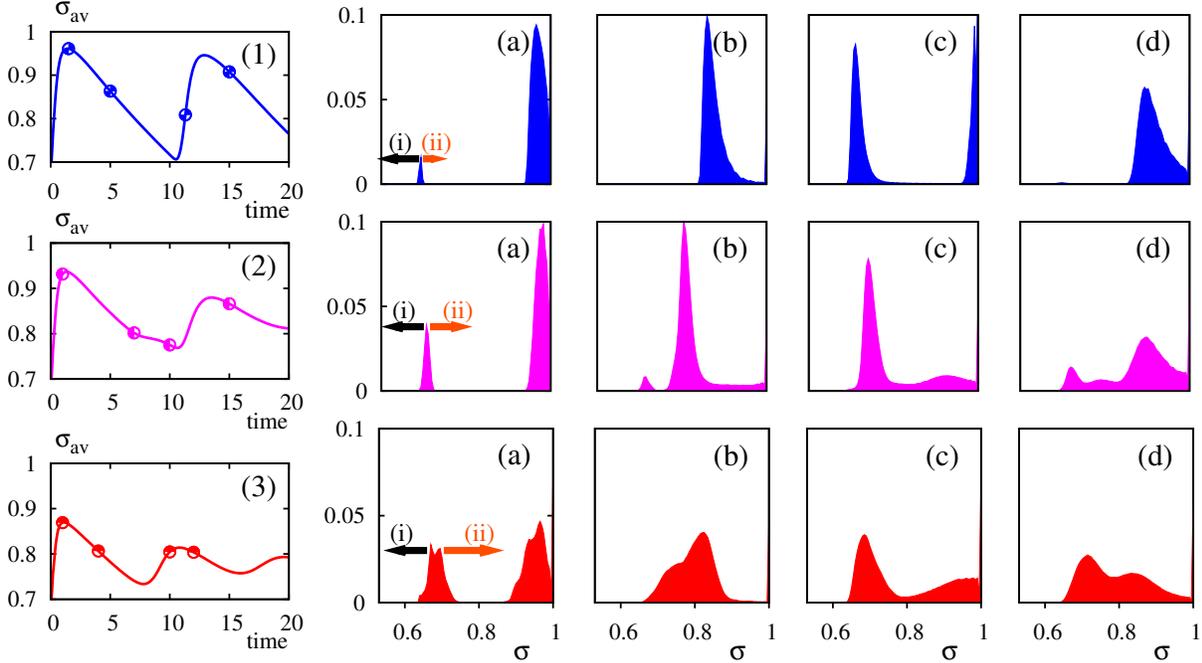} \hfill}
\caption {Evolution of the average composition $\sigma_{\textrm{av}}$ (left panels), and
probability distribution functions of the composition at time instances marked 
by points in the left panels in the weak- ($D=0.05$ -- upper panels), intermediate-
($D=0.15$ -- middle panels), and strong-diffusion regimes ($D=0.55$ -- bottom panels). 
The flow rate is $A=0.8$, the decay rate $\xi=0.04$. The 
simulations were started with a uniform initial composition field
$\sigma_0=0.667$.
The plots of the pdf-s all show the range $0.5 \leq \sigma \leq 1$, the arrows (i) 
and (ii) are described in the text.}
\label{3}
\end{figure*}

\section{Distributions of supersaturation}

At the onset of the nucleation wave the average composition
$\sigma_{av}$ varies in a broad range of values, and does not
always approach the nucleation threshold $\sigma \neq \sigma_{th}$
[see Fig.~\ref{3} panels~(1), (2) and (3)]. This indicates that the onset of
the nucleation wave is not determined by the temporal
evolution of the \emph{average} composition. Rather, the nucleation wave is
initiated by a small fraction of
high-supersaturation ``spots'' that reach the threshold.  Hence, an
adequate description of the phenomenon should deal with the full
distribution of the composition rather than only with the average
composition.  In this section, we present details about the behaviour
of the composition distributions.

We focus on the evolution of the probability distribution functions (pdf-s) of 
the composition field in the low, intermediate- and strong-diffusion regimes. To obtain 
the pdf-s we consider composition bins of size $\delta \sigma =1/200$.
The resulting pdf-s are presented in Fig.~\ref{3} for different time instances:
the upper panels~(1) represent the weak-diffusion regime $D=0.05$, while the middle~(2) and bottom 
panels~(3), the intermediate $D=0.15$, and strong-diffusion $D=0.55$ regimes, respectively.

The initial condition is a uniform composition field
$\sigma(x,0)=0.667$, that is represented in the pdf-s as a
$\delta$-function (not shown). Due to the temperature ramp, this peak moves to the
left, in the direction of the nucleation threshold $\sigma_{th}=2/3$, while
it takes the form of a Gaussian distribution. When few points of the
distribution cross the threshold level the nucleation wave
starts. The points (cells) in which nucleation took place move up to
$\sigma=1$. 
In the  panels (1a), (2a), and (3a) of Fig.~\ref{3} we show the distributions 
at the end of the first nucleation wave. Regardless of the
diffusion regime the behaviour of the $\sigma$-field is similar in
all the three cases: after several cells nucleate droplets the
distribution becomes bimodal. There is a big number of cells being already
around the equilibrium value $\sigma=1$, and a small number of cells
remain still slightly above the threshold and form a second peak.

This small peak is subjected to two opposing forces.  Diffusion, represented 
by the red arrows (ii) in  Fig.~\ref{3}, tends to move the peak to the right: i.e., into the 
direction of the average composition that is already close to equilibrium. The second force is
that of the temperature ramp. It is represented by the black arrows (i) in  Fig.~\ref{3}, and
favours nucleation, i.e., it drives the
peak further to the left in the direction of the nucleation threshold.
The competition between the effects of diffusion and temperature
ramp determines the fate of the small peak. Accordingly, the
character of the oscillations differs in the three diffusion regimes.

In the weak-diffusion regime the effects of diffusion are small 
[(ii)$<$(i) in Fig.~\ref{3} panel~(1a)]. Thus the 
temperature ramp will dominate, and the peak will slowly cross the threshold line
$\sigma_{th}$ and nucleate droplets. This delayed nucleation results in a long 
tail of the main peak, modifying its Gaussian shape [see Fig \ref{3} panel (1b)]. 
The following oscillations
[Fig.~\ref{3} panels (1c) and (1d)] will take place according to the same scenario.
The slowing down is more and more efficient as the diffusion coefficient increases.
Thus, the duration of the nucleation wave increases, explaining the increase of the
oscillation period in the weak diffusion regime. 

In the intermediate diffusion range the effects of the temperature
drift and of diffusion balance each other [(ii)$ \approx $(i) in 
Fig.~\ref{3} panel (2a)]. The small peak is arrested
from nucleation and stopped in the close vicinity of the nucleation
threshold without actually reaching it. Meanwhile, the continuous ramp
of temperature moves the big peak to lower composition levels where
the effect of diffusion becomes less efficient. Eventually, the small
peak manages to cross the nucleation level. This happens {\it after}
the first nucleation wave was finished and gives rise to a separate
small wave of nucleation. In this way, a secondary oscillation of the
droplet density [as shown in Fig.~\ref{1}(b)] occurs in the
system. After this secondary nucleation wave a distinct small peak
forms close to $\sigma=1$ which does not attach to the big peak. Both
peaks maintain their Gaussian character while travelling downward
[Fig.~\ref{3} panel (2b)], maintaining the bimodal distribution of the
composition. In the next wave of nucleation, the bigger peak breaks
again in two distinct modes, forming a trimodal distribution. The
number of modes increases by one at each subsequent principal or
secondary nucleation wave.  Thus, the distribution converges rapidly
to a flat distributions in which the standard deviations of the
individual modes are larger than the distances between them, and the
phase separation proceeds continuously rather than in distinguishable
oscillations.  This explains the rapid decay of the amplitude in this
transition regime.

In the strong-diffusion regime the effects of diffusion overcome the
effects of the temperature ramp [(ii)$>$(i) in 
Fig.~\ref{3} panel (3a)]. The small peak is not only
arrested from nucleation, but it moves backwards towards the big
peak. The big peak in turn, travels downward in the direction of the
nucleation threshold because of the intense diffusive exchange.  
Soon the two peaks merge [Fig.~\ref{3} panel (3b)] and form again one
single peak with an (almost) Gaussian shape [Fig.~\ref{3} panel
(3c)]. In the next nucleation waves everything is repeated in a similar
way [Fig.~\ref{3} panels (3c) and (3d)]. In this regime, there is always a
certain number of cells in which the supersaturation is released by
diffusive transport.
For high values of the diffusion coefficient, the downward travel of
the big peak is accelerated by the rapid diffusive exchange with the
small droplets. It hence moves much faster than in the weak diffusion regime
where this motion was caused by the temperature ramp alone. As a
result the period of the oscillations decreases in this regime.

When starting the simulations with a random initial condition instead
of uniform a $\sigma$-field this picture does not change because diffusion tends
to synchronize the composition. 
In the strong diffusion regime the synchronization 
of lattice cells is so fast that the character (period, amplitude) of the oscillations 
is hardly affected by the initial condition. 
Even in the weak diffusion regime the oscillations are still
clearly observable. Their period remains almost the same, while their amplitude is 
considerably reduced as compared to the case of the uniform initial condition.

\section{Discussion and conclusions}

Oscillatory phase separation of binary mixtures
subjected to slow temperature ramps is a relatively new topic in liquid-liquid 
phase separation.
Even though precision experiments unveiled many characteristics of the
phenomena \cite{VSV97,AVV05,LRVH11}, and there are numerous
theoretical attempts to explain their dynamics
\cite{VVS97,CVWV03,VAV07,V08,BV10}, the parameter dependence of
the oscillations period is still an open question. Secondary effects
appear even in well-controlled experiments \cite[Fig.~15]{LRVH11},
and shield the principal behaviour of the system. For instance, 
\textbf{(a)}~secondary oscillations can appear \cite{Tobias},
\textbf{(b)}~the frequency of the oscillations increases during one 
experimental run (Fig.~10 in \cite{AVV05};  Fig.~3 in \cite{VAV07}; Fig.~15 in \cite{LRVH11}), 
and \textbf{(c)}~bimodal droplet size distributions occur \cite[Fig.~3]{S03}.

In this communication we identified these secondary effects in a model system,
and we pointed out that they can be 
attributed to the growing diffusion coefficient
as the system departs from the critical point.
Upon increasing the diffusion coefficient we observe a cross-over regime where 
the behaviour of the demixing changes qualitatively.

\textbf{(a)} In the cross-over regime the system performs a secondary
oscillation between any two major bursts of nucleation
[Fig.~\ref{1}(b)].  Similar secondary oscillations are clearly visible
in experiments when the system is still close to the critical point,
and accordingly the diffusion coefficient is relatively small
\cite{Tobias}. The secondary oscillations are observable over several
periods. In the experiments they disappear eventually, as the
diffusion coefficient increases due to the change of temperature.

\textbf{(b)} Our result, Fig.~\ref{2}(a), shows that in experimental
runs where the diffusion coefficient increases gradually as the system
departs from the critical point, one should expect that the period of
the oscillations might initially increase slightly and eventually
decrease considerably.  This has indeed been observed in experiments
(cf.~Fig.~10 in \cite{AVV05}; Fig.~3 in \cite{VAV07}; Fig.~15 in
\cite{LRVH11}).

\textbf{(c)} Exploring the distribution of the supersaturation
has revealed that the oscillatory phase separation can not be described
adequately in terms of the average composition.  The spatial
distributions of the composition plays a crucial role, and should
explicitly be taken into account.  This is particularly important in
the cross-over regime where the secondary oscillations are
observed. They are consequences of the bimodal distributions that
temporarily appear in the evolution of the composition field.
Accordingly, droplets are formed at two different time instances
during an oscillation. Since droplets start to grow immediately after
they are nucleated, this difference in the nucleation time will lead
to bimodal droplet size distributions as they have been observed in
\cite[Fig.~3]{S03}. We expect that such bimodal droplet size
distributions appear only in the intermediate and strong-diffusion
regimes.  An experimental test of these predictions is under way
\cite{Tobias}.

In summary, we conclude that identifying secondary effects which arise
from the interplay of diffusion and the non-trivial spatial
distribution of supersaturation provides an important step to a
comprehensive understanding of the dynamics of phase separation in the
presence of a temperature ramp.

\acknowledgements

We are grateful to Tobias Lapp, Martin Rohloff, Ray Pierrehumbert, and Tam\'as T\'el for
discussions, inspiration and feedback.

\end{document}